# Analysis of random telegraph noise in resistive memories: The case of unstable filaments

Nikolaos Vasileiadis [a,b,*], Alexandros Mavropoulis [a], Panagiotis Loukas [a], Georgios Ch. Sirakoulis [b], Panagiotis Dimitrakis [a]

[a] *Institute of Nanoscience and Nanotechnology, NCSR "Demokritos", Ag. Paraskevi 15341, Greece*
[b] *Department of Electrical and Computer Engineering, Democritus University of Thrace, Xanthi 67100, Greece*

## ABSTRACT

Through Random Telegraph Noise (RTN) analysis, valuable information can be provided about the role of defect traps in fine tuning and reading of the state of a nanoelectronic device. However, time domain analysis tech-niques exhibit their limitations in case where unstable RTN signals occur. These instabilities are a common issue in Multi-Level Cells (MLC) of resistive memories (ReRAM), when the tunning protocol fails to find a perfectly stable resistance state, which in turn brings fluctuations to the RTN signal especially in long time measurements and cause severe errors in the estimation of the distribution of time constants of the observed telegraphic events, i.e., capture/emission of carriers from traps. In this work, we analyze the case of the unstable filaments in silicon nitride-based ReRAM devices and propose an adaptive filter implementing a moving-average detrending method in order to flatten unstable RTN signals and increase sufficiently the accuracy of the conducted measurements. The $\tau_e$ and $\tau_c$ emission/capture time constants of the traps, respectively, are then calculated and a cross-validation through frequency domain analysis (Lorentzian fitting) was performed proving that the proposed method is accurate.

## 1. Introduction

Resistive memories (ReRAMs) are one of the most anticipated non-volatile memory technologies [1]. The crossbar architecture offers an excellent opportunity for achieving the smallest memory cell [2], a building block for in-memory, neuromorphic and quantum computing accelerators [3,4], and the ensuing development in low power mobile IoT edge computing hardware [5,6]. ReRAM cells are ideal for this ar-chitecture [6,7]. However, there are a lot of challenges until commer-cialization in all fabrication levels also including the physics behind the resistive switching phenomenon [8].

The analysis of low-frequency noise (LFN) signals generated by semiconductor devices yields important knowledge about their physical mechanisms and origin, commonly related to the presence of traps [9], enabling the reduction of noise sources and enhancing device perfor-mance, respectively [10]. More specifically, random telegraph noise (RTN) analysis [9,11] can provide important information about the energy of the traps and their mobility thus playing an important role in the fine tuning of the nanoelectronic device [9]. Furthermore, multi-level resistance tuning is a fundamental desired aspect on all ReRAM devices to be used in in-memory and neuromorphic computing [12]. The origin of the RTN signal in filamentary RRAM devices is mainly attrib-uted to the exchange of electrons among the conductive filamentary paths and the traps of the surrounding dielectric [13,14]. Also, a model for the estimation of trap location and energy in $TiO_x$ RRAM cells was proposed [15]. Nevertheless, it has been observed that multiple resis-tance states are not stable in time and a trend (instability) appears in the RTN signal. Several such RTN signals have been reported for MOSFET [16,17] as well as for RRAM devices [18]. Thus, the analysis of such signals in time domain becomes a challenging task.

In the present research, an adaptive filter that implements a moving-average detrending algorithm is proposed to flatten the unstable RTN signals and, moreover, make them suitable for further time domain analysis. The filter takes into account key parameters of the measured unstable RTN signals that needs to be stabilized – such as mean and standard deviation values of the Gaussian lobes in RTN signal's histo-gram – and performs an optimization loop algorithm in order to find the best averaging filter mask size that minimizes a custom criterion, which is composed by these parameters. It is also worth to mention that the proposed adaptive filter, with some modifications, is possible to be adapted as a stabilizing mechanism between the entropy source and the harvesting entropy circuit aiming to reduce the need of re-tuning the ReRAM device and lower the overall energy consumption of the circuit, a completely necessary element for IoT applications [5,6].

Despite the fact that this problem was addressed for the first time in the present work, to the best of our knowledge, with a more systematic approach that also considers frequency domain parameters a corresponding article [19] has been published recently in a quite different field but with a similar conception. In particular, Cheveigné et al. applied a robust detrending method on corrupted electroencephalography measurements and they managed not to discard but to reuse all the corrupted measurements with all the desired information to be included and at the same time to remove the unwanted artifacts that a simple filtering method would produce. Furthermore, in a recent publication [20] different algorithms for the statistical analysis of RTN signal were compared using simulated RTN signals. Among the Baum-Welch and the measurement discretization algorithms, the Hidden Markov Model (HMM) was shown that can perform slightly better the monitor of the current deviations in order to extract the RTN statistical parameters. Nevertheless, the case of unstable filament and corresponding RTN signals is not considered in that article. Such a case is presented in [21] where unstable RTN signal, in HfOx ReRAM devices, were analyzed with HMM algorithm. However, the results are not so encouraging when experimental unstable filament were modeled.

In the proposed herein approach, experimental data obtained from silicon nitride resistive memory cells are used to stress the efficiency of the introduced method. It should be mentioned that the origin of the RTN signal in such ReRAM nanodevices is due to the interaction of electrons, flowing in silicon nitride through conductive filaments, with the intrinsic nitride defects surrounding the filaments. The mechanism has been described and discussed in detail in literature [9,22–24].

The rest of the paper is organized as following. In Section 2, the possible origin of unstable signals in ReRAM devices are briefly summarized. Next, the approach of the proposed adaptive filter method is presented analytically. In Section 3, the results of the method after its application to real measured RTN unstable signals are introduced and thoroughly discussed. Finally, Section 4 concludes the paper.

## 2. Challenge definition and counteract approach

### 2.1. Instabilities in silicon nitride ReRAM read currents

The most probable mechanism for bipolar operation of silicon nitride ReRAM is the formation of conductive filaments (CF) either due to the coordination of intrinsic defects at high-electric field areas or due to the nitrogen vacancies (NV) as in valence change memories (VCM) [22]. In this filamentary kind of resistive memory cells, the observed read cur-rent instabilities in time are due to unstable filaments. The conductive filaments short-circuit the two electrodes, Cu and $n^{++}$-Si, having a 7 nm $SiN_x$ switching layer in between, by applying a positive voltage on Cu top-electrode (TE). In this case, the resistance of SiN layer is low and becomes high when the nanometer-sized filament is disrupted by applying a negative voltage on TE. The formation and the disruption of CFs is a stochastic process as well as their geometrical characteristics, like the length and the diameter [25]. The presence of RTN is the present $SiN_x$ RRAM MIS devices have been already investigated and demon-strated [9,22–24]. Usually, two-level RTN signal is observed corre-sponding to the electron capture and emission from single-trap close to the CF. In valence change RRAM devices, like the ones examined here, the CF filaments have semiconductor properties. Thus, the trap should be at a distance of one Debye length [13].

In Fig. 1, unstable read current timeseries, obtained from silicon nitride ReRAM cells, are presented. Unstable filaments are observed either in high-resistance state (HRS) or in low-resistance state (LRS) (Fig. 1(a)). For some cells, the measured read current instabilities include upward and downward movements, which are mainly attributed to CF resistance instabilities caused by the variation of the number of defects/vacancies comprising the CF. Primarily, the variation of defects/ vacancies in the CF as well as their excess non-uniform generation is governed by thermal effects inside the switching layer [26]. Additional effects that contribute to resistance temporal instability are the number of parallel CFs or a single CF consisted by a bundle of conductive paths (Fig. 1(b)). Resistance retention failure is characterized by a non- reversible [27] and/or monotonic current decrease (LRS) or increase (HRS) [28,29] caused by permanent filament wear out (Fig. 1(c)).

A plethora of models describing the possible mechanisms responsible for unstable RTN signals have been proposed. Among them, metastable defects [30] and electrostatic interactions between defects [31] could explain filament instabilities causing unstable RTN signal. Specifically, under charging/discharging conditions the traps intermediate configuration modes due to the different coordination of the defect atoms, leading to metastable states having usually faster time constants. In this case, additional current levels are observed in RTN signal [30]. Furthermore, RTN complexity increases when the traps can interact electrostatically. In such a case, the local electrostatic field around a defect is modified according to the charging state of the trap, especially when the applied bias during the RTN measurement is small. It is important to remark that the change of the charging state of traps surrounding the CF due to common processes, could affect electrostat­ically the potential in the CF introducing additional number of current levels in the recorded I-t noise signal. Such an electrostatic interaction can have positive or negative contribution to the current depending on the coupling model (series or parallel) and the charging state of the traps [35]. The electrostatic interaction modifies the trap time constants as well as introduces more states in the RTN signal [31].

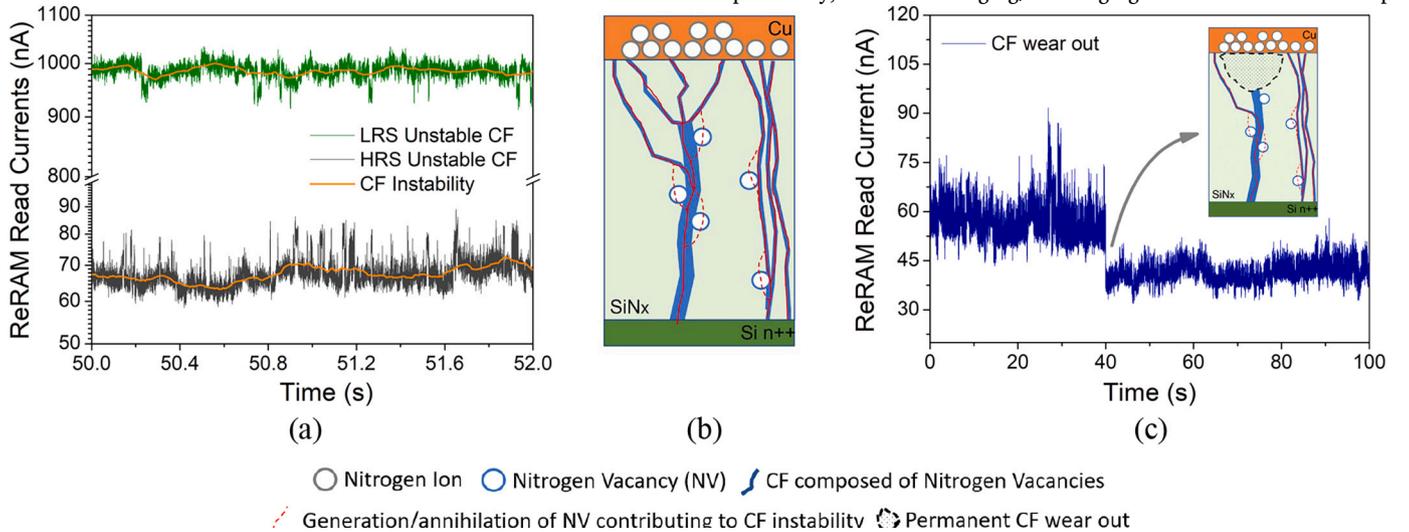

**Fig. 1.** (a) HRS and LRS read currents of silicon nitride ReRAM device that appear RTN along with conductive filament instabilities, (b) a schematic representation of possible conduction instability mechanisms such as generation/annihilation of nitrogen vacancies, variation of the number of parallel CFs or the number of CF in a bundle, and (c) an example of HRS retention failure due to permanent CF wear out.

### 2.2. Proposed methodology

The signal trending or instability should be removed before any further LFN signal analysis, e.g., the calculation of the capture time $\tau_c$ and emission time $\tau_e$ constants of carrier traps. Thus, a dedicated methodology is requested to tackle efficiently the problem, which is the exact scope of this work. Assuming the existence of a single-trap, two distinct levels exist in the measured signal. In order to analyze the method of adaptive filtering, we will start from a decomposition of the problem.

In Fig. 2(a), a simulated RTN signal is shown when it is pure $1/f^2$ noise (upper figure) and after a flicker noise signal is added (lower figure). The latter signal is generated by inverse Fast Fourier Transform (FFT) of a simulated $1/f$ power spectral density (PSD) function. The pure RTN signal was created by two pseudorandom processes (Fig. 2(d)) which follow the exponential distribution of Eq. (1):

$$g(t) = \frac{1}{\tau} e^{-\frac{t}{\tau}} \tag{1}$$

where $t$ is the time and $\tau$ is the effective time constant:

$$\frac{1}{\tau} = \frac{1}{\tau_e} + \frac{1}{\tau_c} \tag{2}$$

The time constants $\tau_c$ and $\tau_e$ which were chosen in each case together with the plots of their exponential distributions are shown in Fig. 2(d). In order to be thorough, $\tau_c$ represents the time elapsed between two suc­cessive trapping events and $\tau_e$ represent the time elapsed between two successive emission events of a trapped carrier. Recently, Kinetic Monte Carlo algorithm was used to simulate RTN signals in RRAMs [30].

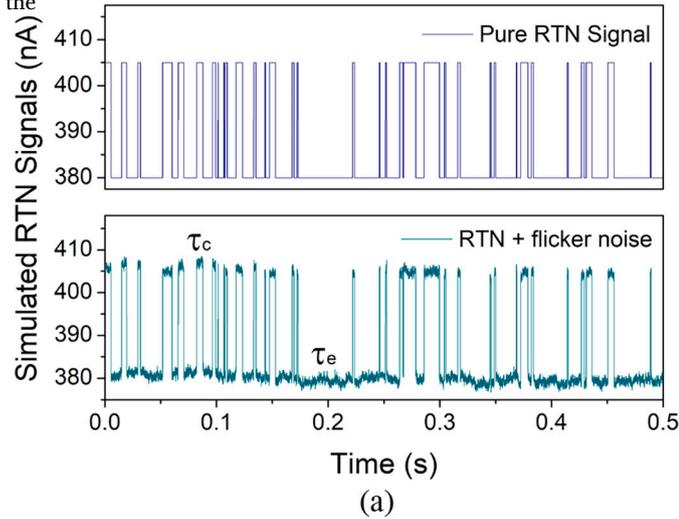

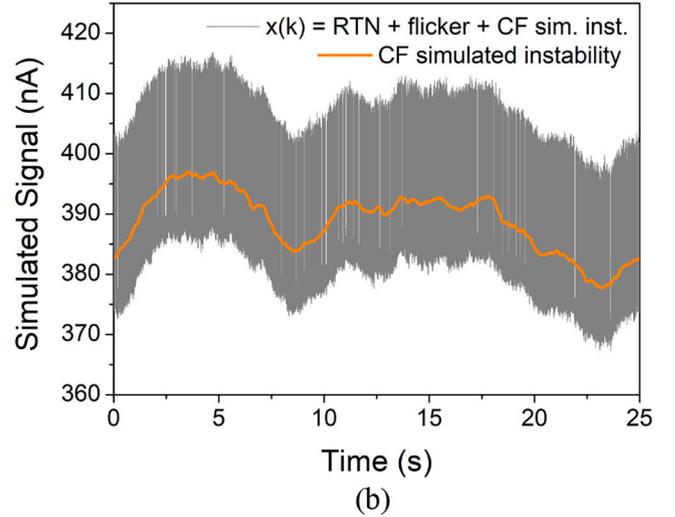

(a)

(b)

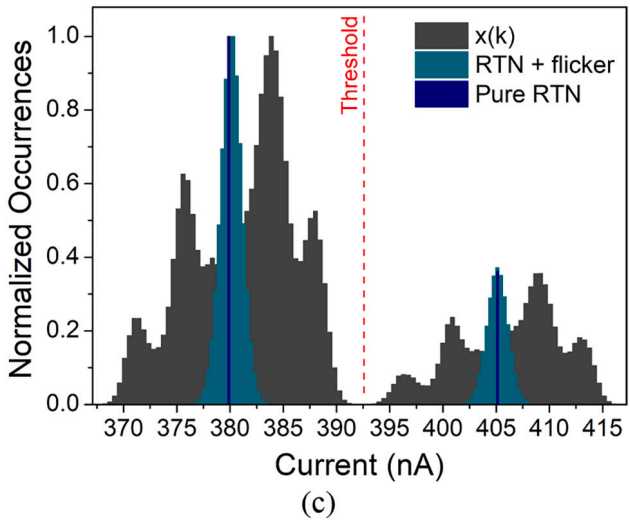

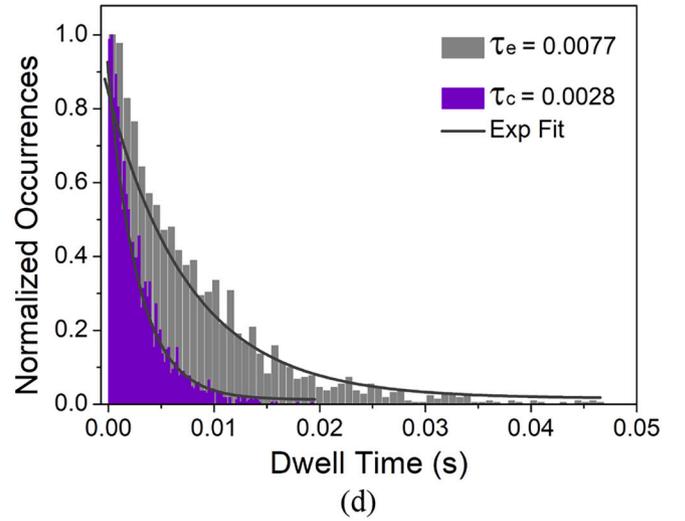

(c)

(d)

**Fig. 2.** The characteristics of the simulated RTN signal with CF instability, which was considered during the development of introduced the methodology as presented in this section. Figure (a) includes a pure $1/f^2$ RTN signal (upper figure) and a $1/f$ noise signal over pure RTN (lower figure), (b) includes the simulated unstable RTN signal, (c) includes the histograms of the previous signals and (d) shows the exponential distributions of the stochastic processes which were used in order to produce the pure RTN signal.

In Fig. 2(c), the histograms of the two signals of Fig. 2(a) are shown in comparison aiming to make clear the effect of the flicker noise when it is added to the pure RTN signal. Up to this point, the effect is not total distractive and the two levels of the RTN signal can be separated still easily with a simple threshold between the two Gaussian lobes. The signal processing is getting more complicated when the "instability signal" also makes its appearance due to slowly varying disturbances. Then our signal takes the form of Fig. 2(b). As can be seen from the histogram of this signal (Fig. 2(c)), it becomes more and more difficult to determine how many levels there are in the RTN signal and subsequently to extract correctly the average time constants $\tau_e$ and $\tau_c$. Such a signal instability can be simulated with many different approaches. In this example a $1/f^3$ signal was used with an order of magnitude smaller corner frequency $f_0$' than the simulated RTN signal's corner frequency $f_0$. We may state that as a general principle for producing such an instability is that $f_0$' should be at least an order of magnitude lower than the $f_0$ and also the power spectral density of the simulated instability signal should be at least 3 orders of magnitude smaller than that of the simulated RTN signal, in order to prevent any increased interference from making further processing impossible with the method we are analyzing. For the examined SiN$_x$ RRAM devices, the dominant traps are the nitrogen vacancies which act as donor traps and so they are naturally positive charges. So, the average times of high- and low-current levels correspond to the emission and capture times, respectively [15].

To solve this problem, an adaptive filtering method is proposed hereafter and presented in Fig. 3 accordingly. The core of this filter is a moving average detrending algorithm, which is based on relation (3):

$y_M(k) = x(k) - (x*G_M)(k) + \overline{x}(k)$ with

$$G_M(k) = \left[\underbrace{1/M, 1/M, \ldots, 1/M}_{M\ times}\right] \quad (3)$$

where $x(k)$ is the unstable RTN signal and $G(k)$ the size $M$ mean filtering mask. Firstly, the $x(k)$ signal is filtered with a mean filter and the resulting signal is subtracted from the $x(k)$ signal, while also the average value of the $x(k)$ signal, $\overline{x}(k)$, is added, so as to preserve an estimation of the original DC component of the RTN. Our goal is to find the appropriate size of the filtering mask in order to have the minimum average error between the simulated RTN signal and the $y(k)$ signal. The reason why a mean filter instead of a low pass filter was chosen is twofold. Firstly, the resulting simpler implementation and secondly, the requested computational time that becomes important during the iterative process of adaptive filtering.

Fig. 4(a) shows the change of the aforementioned error in relation to the dimension of the mask size used in mean filtering of the simulated $x(k)$ signal depicted in Fig. 2(b). A global minimum is evident.

Nevertheless, in a real case scenario, the ground truth RTN signal is unknown and thus the absolute error. As a result, another minimization criterion should be found, which will drive the proposed adaptive filter minimization algorithm. Such a criterion is defined by the minimization of following quantity named $C$:

$$C = a \bullet \frac{(\sigma_c + \sigma_e)}{|\mu_c - \mu_e|} + b \bullet \left[\left|f_0 - \frac{1}{2\pi}\left(\frac{1}{\tau_c} + \frac{1}{\tau_e}\right)\right| \Big/ N\right] = a \bullet A + b \bullet B \quad (4)$$

where $\sigma$ is the standard deviation, $\mu$ is the mean value, while $a$ and $b$ are scaling parameters, $f_0$ the PSD's corner frequency, $\tau_e$ the emission time constant, $\tau_c$ the capture time constant, respectively and $N$ a quantization factor. As described in Fig. 3, the adaptive filter recalculates and feeds the minimization algorithm with these 6 parameters, namely ($\mu_c, \mu_e, \sigma_c, \sigma_e, \tau_c, \tau_e$), after each iteration of the moving average detrending al-gorithm. The parameter $f_0$ needs to be calculated only once from the PSD of the $x(k)$ signal and refers to the corner frequency of the RTN (Fig. 4 (b)) which meets well the introduced criteria of eq. (5) when it comes to single trap RTN [32].

$$f_0 = \frac{1}{2\pi}\left(\frac{1}{\tau_c} + \frac{1}{\tau_e}\right) \quad (5)$$

The $C$ criterion of eq. (4) resulted from the observation of the changes of ($\mu_c, \mu_e, \sigma_c, \sigma_e, \tau_c, \tau_e$) parameters in respect of mean filtering mask size. More specifically, as shown in Fig. 4(d), for a small size mean filtering mask the mean values of the two Gaussian lobes of the histo-gram of the detrended RTN signal tend to get closer while for large ones they drift apart. On the contrary, the variation of the current values around the mean values tends to increase for large and small mean filter mask sizes but to decrease for medium ones, as shown in Fig. 4(e). This happens because with small size masks, the adaptive filter tends to overestimate unstable RTN fluctuations and after the detrending this is attributed to the histogram with greater standard deviation, as shown in Fig. 4(g). In the case of large size masks, we do not manage to remove all the instabilities of the $x(k)$ signal and for this reason secondary peaks are observed in the histogram, which in turn increase the standard devia-tion, as presented in Fig. 4(i). Where the standard deviations are mini-mized, we also receive the optimal solution (Fig. 4(h)). All this information is included in term $A$ of the $C$ criterion.

In order to reduce the searching space area for the minimization algorithm evenmore, an extra term $B$ in the criterion's $C$ eq. (4) has been used. Now, the algorithm will try to minimize only near corner fre-quency $f_0$. We have also included a quantization parameter $N$ to reduce the fluctuations of term $B$, which are caused due to small fitting error that is observed during the $\tau_c$ and $\tau_e$ parameter estimation through dwell time histograms exponential fitting. This phenomenon intensifies the more noise found in our data and in such cases, it and is recommended to use a more complex algorithm for calculating the levels of the estimated RTN signal than a simple threshold-based levels detection algorithm [33]. Fig. 4(c) shows how the $N$ parameter creates a "valley" and also smooths out the fluctuations feeding the minimization algorithm with a smoother curve in order to improve the speed of convergence.

The parameters a and b are scaling parameters and were placed in order to give a more generic form to the equation, while parameter $N$ is the integer term with which the integer division-quantization of term $B$ is done. These parameters give greater flexibility to the user in order to build a smoother error surface through which the minimization algorithm will converge to the desired solution. The user should run the adaptive filter algorithm as many times as necessary to the estimate parameters $\tau_c$ and $\tau_e$ with the smallest possible deviation from $f_0$ through the relation (5). A rule of thumb for the *a* and *b* parameters is to be near 1 while *N* parameter near $f_0$. In an improved future version of the adaptive filter, estimation could be done for these parameters as well, however the performance of the algorithm should be reconsidered.

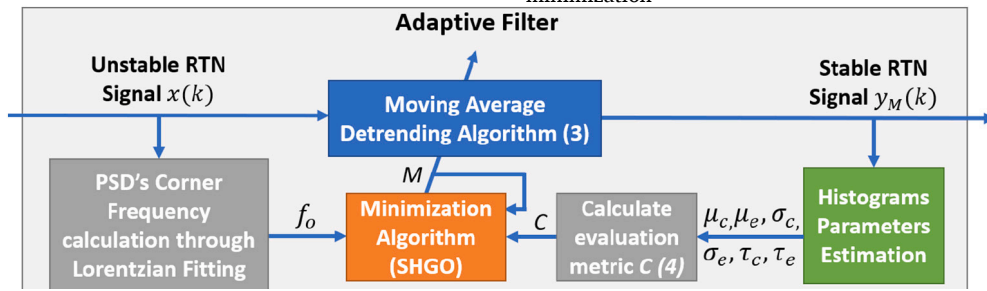

**Fig. 3.** A representation of the structure of the proposed adaptive filter.

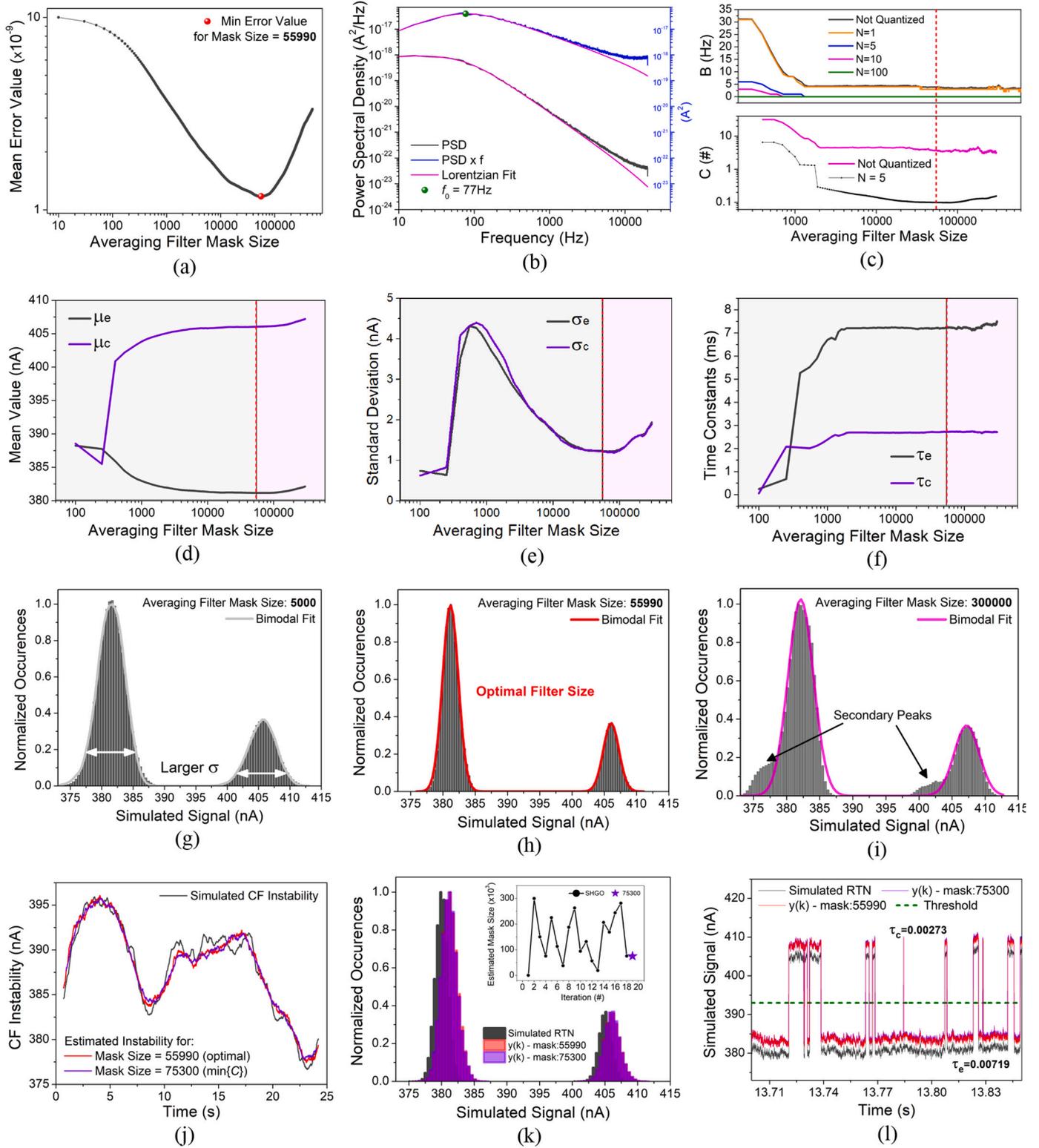

**Fig. 4.** Auxiliary graphs in order to help the analysis of the proposed adaptive filtering methodology. Plot (a) shows the variation of the mean | x(k) – y(k) | in relation to the size of the mask of the averaging filter that was used for the estimation, (b) shows the PSD of the simulated RTN signal along with a Lorentzian fit estimation of the corner frequency $f_0$, (c) shows the criterion $C$ changes by varying key parameters, while (d), (e), (f) show accordingly the variations of $\mu$, $\sigma$ and $\tau$ parameters of the estimated RTN y(k) respectively, in relation to the size of the mask of the averaging filter that was used for the estimation (g), (h), (i) show histogram distortions for below optimal, optimal (not distorted) and above optimal averaging filter mask size respectively (j), (k) and (l) show the simulated-estimated CF instability, RTN histogram and RTN signal respectively, for different averaging filter mask sizes (the inset shows the minimizations algorithm iteration steps).

Fig. 4(j) shows the result of the proposed adaptive filtering on the signal of Fig. 2(b), regarding the estimation of the injected instability signal. The minimization was done through the simplicial homology global optimisation (SHGO) algorithm [34], the iterations of which are shown in the inset of Fig. 4(k). The SHGO algorithm was chosen because it is a global optimizer which also avoids local minima. It is slower from a local optimizer but definitely faster than a brute force approach, which in this case the accurate estimation of the timing parameters $\tau_c$ and $\tau_e$ of the RTN signal is more important than real-time program execution, while narrowing the searching space with the criterion C's quantization parameter N of eq. (3) can drastically increase the performance. Furthermore, in Fig. 4(k), the estimated RTN $y(k)$ histogram after removing the instability signal is shown and compared with the ground truth RTN signal. Fig. 4(l) shows the adaptive filter output. This is the time frame with the biggest deviation between the simulated and the estimated RTN signal, but still the two levels in the RTN signal can be separated with a threshold and their time constants $\tau_e$ and $\tau_c$ can be calculated accurately.

## 3. Results and discussion

In this section, the application of the proposed methodology which was described previously is utilized to analyze the RTN signal measured on a silicon nitride ReRAM device that has been tuned in an intermediate resistance level. Fig. 5(a) shows such a signal and the corresponding PSD is presented in Fig. 5(b). The corner frequency is found $f_0 = 985$ Hz after Lorentzian fitting. In order to estimate $\tau_c$ and $\tau_e$ constants, the adaptive filter algorithm is executed. Fig. 5(e) shows the search space specified by the criterion (4) for different averaging filter mask sizes and quantiza-tion constant $N = 220$. Our algorithm continues to perform well and the confirmation criterion -relation (5)- is satisfied for the estimated $\tau_c$ and $\tau_e$ parameters, as clearly shown in Fig. 5(f). Obviously, from the time frame shown in inset of Fig. 5(c), the method of extracting the time constants fails completely without the adaptive filtering, since the un-stable signal is above the threshold. This becomes even clearer in Fig. 5 (d) where the two Gaussian lobes are mixed together before adaptive filtering.

## 4. Conclusions

In this work, an extensive presentation of a method to flatten unstable RTN signals consisting of a single trap was presented. The reasons that cause instabilities in silicon nitride ReRAM devices were summarized providing a guideline about which signals could be treated following the introduced method. An adaptive filter with a robust minimization criterion was efficiently designed. The effectiveness of the method was demonstrated through its successful implementation on a real measured unstable RTN signal of silicon nitride ReRAM nanodevice. For future work, the case of unstable RTN signals with multiple traps will be analyzed, as well as an optimization of this method in order to be more easily incorporated in IoT TRNG security systems.

**Declaration of Competing Interest**

The authors declare the following financial interests/personal

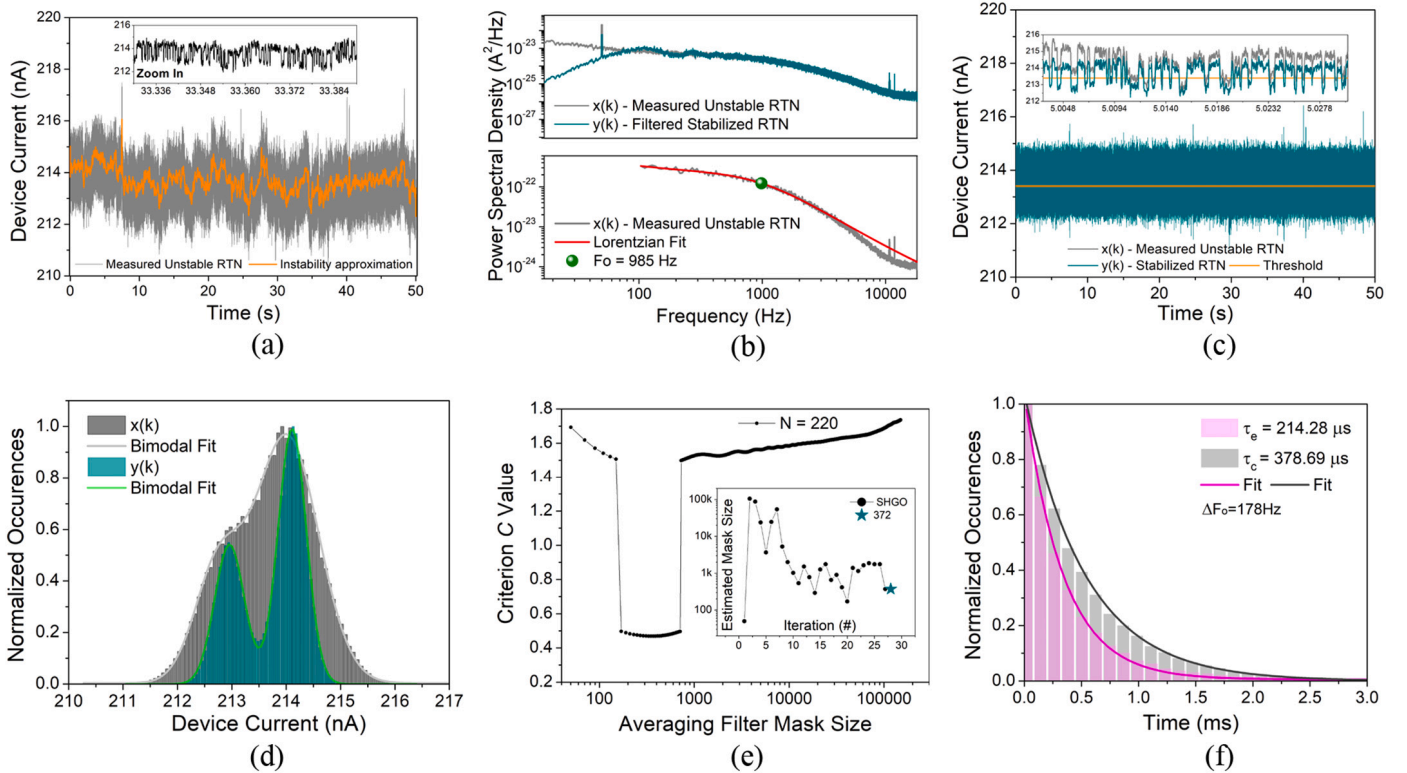

**Fig. 5.** Experimental results of the proposed adaptive filtering method applied on a measured unstable RTN signal. Plot (a) shows the corresponding measured unstable RTN signal, while (b) presents the spectral differences of the measured unstable and the stabilized RTN signal. Lower figure of (b) validates that all subtracted frequencies were below corner frequency $f_0$, (c) shows the stabilized RTN signal, (d) depicts unstable versus stabilized RTN signal's histogram, (e) introduces the c-criterion shaped solution search space (the inset shows the minimizations algorithm iteration steps) for different mask sizes, and (f) shows the time constants and the dwell time distributions for the estimated levels of the stabilized RTN signal.

relationships which may be considered as potential competing interests:

N. Vasileiadis reports financial support was provided by Hellenic Foundation for Research and Innovation. P. Dimitrakis reports financial support was provided by Hellenic Foundation for Research and Innovation. A. Mavropoulis reports financial support was provided by General Secretariat for Research and Technology. P. Loukas reports financial support was provided by General Secretariat for Research and Technology.

**Data availability**

Data will be made available on request.

**Acknowledgments**

This work was supported in part by the research projects "3D-TOPOS" (MIS 5131411) and "LIMA-chip" (Proj.No. 2748) which are funded by the Operational Programme NSRF 2014-2020 and the Hellenic Foundation of Research and Innovation (HFRI) respectively.

**References**

[1] Ch. Yangyin, ReRAM: history, status, and future, IEEE Trans. Elect. Dev. 67 (4) (2020) 1420–1433, https://doi.org/10.1109/TED.2019.2961505.

[2] A.H. Edwards, H.J. Barnaby, K.A. Campbell, M.N. Kozicki, W. Liu, M.J. Marinella, Reconfigurable memristive device technologies, Proc. IEEE 103 (7) (2015) 1004–1033, https://doi.org/10.1109/JPROC.2015.2441752.

[3] S. Mittal, A survey of ReRAM-based architectures for processing-in-memory and neural networks, Machine Learn. Knowledge Extract. 1 (1) (2018) 75–114, https://doi.org/10.3390/make1010005.

[4] I.-A. Fyrigos, et al., Memristor crossbar arrays performing quantum algorithms, IEEE Trans. Circuits Syst. I: Regular Papers 69 (2) (Feb. 2022) 552–563, https://doi.org/10.1109/TCSI.2021.3123575.

[5] N. Vasileiadis, V. Ntinas, P. Karakolis, P. Dimitrakis, G. Ch. Sirakoulis, On edge image processing acceleration with low power neuro-memristive segmented crossbar array architecture, Int. J. Unconv. Comput., 17(3) 173–199.

[6] A. Shamsoshoara, A. Korenda, F. Afghah, S. Zeadally, A survey on physical unclonable function (PUF)-based security solutions for internet of things, Comput. Netw. 183 (2020) 107593, https://doi.org/10.1016/j.comnet.2020.107593.

[7] N. Vasileiadis, P. Dimitrakis, V. Ntinas, G. Ch, Sirakoulis, true random number generator based on multi-state silicon nitride memristor entropy sources combination, in: 2021 International Conference on Electronics, Information, and Communication (ICEIC), IEEE, 2021, https://doi.org/10.1109/ICEIC51217.2021.9369817.

[8] Z. Wang, H. Wu, G.W. Burr, C.-S. Hwang, K.L. Wang, Q. Xia, J.J. Yang, Resistive switching materials for information processing, Nat. Rev. Mater. 5 (3) (2020) 173–195, https://doi.org/10.1038/s41578-019-0159-3.

[9] N. Vasileiadis, P. Loukas, P. Karakolis, V. Ioannou-Sougleridis, P. Normand, V. Ntinas, I.-A. Fyrigos, I. Karafyllidis, G.Ch. Sirakoulis, P. Dimitrakis, Multi-level resistance switching and random telegraph noise analysis of nitride based memristors, Chaos, Solitons Fract. 153 (1) (2021), 111533, https://doi.org/10.1016/j.chaos.2021.111533.

[10] V. Ntinas, et al., Power-efficient noise-induced reduction of ReRAM cell's temporal variability effects, IEEE Trans. Circ. Syst. II: Express Briefs 68 (4) (2020) 1378–1382, https://doi.org/10.1109/TCSII.2020.3026950.

[11] S. Ambrogio, S. Balatti, A. Cubeta, A. Calderoni, N. Ramaswamy, D. Ielmini, Understanding switching variability and random telegraph noise in resistive RAM, in: 2013 IEEE International Electron Devices Meeting, IEEE, 2013, https://doi.org/10.1109/IEDM.2013.6724732.

[12] Chaohan Wang, et al., Multi-state Memristors and their applications: an overview, IEEE J. Emerg. Select. Topics Circuits Systems 12 (4) (2022) 723–734, https://doi.org/10.1109/JETCAS.2022.3223295.

[13] Daniele Ielmini, Federico Nardi, Carlo Cagli, Resistance-dependent amplitude of random telegraph-signal noise in resistive switching memories, Appl. Phys. Lett. 96 (2010), 053503, https://doi.org/10.1063/1.3304167.

[14] D. Veksler, et al., Random telegraph noise (RTN), in: In Scaled RRAM Devices, 2013 IEEE International Reliability Physics Symposium (IRPS), Monterey, CA, USA, 2013, https://doi.org/10.1109/IRPS.2013.6532101 pp. MY.10.1-MY.10.4.

[15] Jung-Kyu Lee, et al., Extraction of trap location and energy from random telegraph noise in amorphous TiOx resistance random access memories, Appl. Phys. Lett. 98 (2011), 143502, https://doi.org/10.1063/1.3575572.

[16] Eddy Simoen, et al., Random telegraph Noise: from a device Physicist's dream to a Designer's nightmare, ECS Trans. 39 (2011) 3.

[17] R. Wang, S. Guo, Z. Zhang, J. Zou, D. Mao, R. Huang, Complex random telegraph noise (RTN): What do we understand?, in: 2018 IEEE International Symposium on the Physical and Failure Analysis of Integrated Circuits (IPFA), Singapore, 2018, pp. 1–7, https://doi.org/10.1109/IPFA.2018.8452514.

[18] Francesco Maria Puglisi, Luca Larcher, Andrea Padovani, Paolo Pavan, Anomalous random telegraph noise and temporary phenomena in resistive random access memory, Solid State Electron. 125 (2016) 204–213, https://doi.org/10.1016/j.sse.2016.07.019.

[19] A. de Cheveigné, D. Arzounian, Robust detrending, rereferencing, outlier detection, and inpainting for multichannel data, NeuroImage 172 (2018) 903–912.

[20] Cláudia Theis da Silveira, et al., Implementation and comparison of algorithms for the extraction of RTN parameters, in: 2021 35th Symposium on Microelectronics Technology and Devices (SBMicro), IEEE, 2021.

[21] Francesco M. Puglisi, et al., Random telegraph signal noise properties of HfOx RRAM in high resistive state, in: 2012 Proceedings of the European Solid-State Device Research Conference (ESSDERC), IEEE, 2012.

[22] N. Vasileiadis, P. Loukas, A. Mavropoulis, P. Normand, I. Karafyllidis, G. Ch. Sirakoulis, P. Dimitrakis, Random Telegraph Noise of MIS and MIOS Silicon Nitride memristors at different resistance states, in: 2022 IEEE 22nd International Conference on Nanotechnology (NANO), 2022, pp. 453–456, https://doi.org/10.1109/NANO54668.2022.9928707.

[23] N. Vasileiadis, A. Mavropoulis, P. Loukas, P. Normand, G.C. Sirakoulis, P. Dimitrakis, Substrate Effect on Low-frequency Noise of synaptic RRAM devices, in: 2022 IFIP/IEEE 30th International Conference on Very Large Scale Integration (VLSI-SoC), 2022, pp. 1–5, doi: 0.1109/VLSI-SoC54400.2022.9939652.

[24] A. Mavropoulis, N. Vasileiadis, C. Theodorou, L. Sygellou, P. Normand, G. Ch. Sirakoulis, P. Dimitrakis, Effect of SOI substrate on silicon nitride resistance switching using MIS structure, Solid State Electron. 194 (2022), 108375, https://doi.org/10.1016/j.sse.2022.108375.

[25] Daniele Ielmini, Rainer Waser (Eds.), Resistive Switching: From Fundamentals of Nanoionic Redox Processes to Memristive Device Applications, John Wiley and Sons, 2016.

[26] F.T. Chen, et al., Resistance instabilities in a filament-based resistive memory, in: 2013 IEEE International Reliability Physics Symposium (IRPS), 2013, https://doi.org/10.1109/IRPS.2013.6532040, pp. 5E.1.1–5E.1.7.

[27] D. Veksler, et al., Methodology for the statistical evaluation of the effect of random telegraph noise (RTN) on RRAM characteristics, in: 2012 International Electron Devices Meeting, San Francisco, CA, USA, 2012, https://doi.org/10.1109/IEDM.2012.6479013, pp. 9.6.1-9.6.4.

[28] A. Ghetti, C.M. Compagnoni, A.S. Spinelli, A. Visconti, IEEE Transac. Electron. Devices 56, 2009, p. 1746.

[29] C. Wen, et al., Advanced data encryption using 2D materials, Adv. Mater. 33 (2021) 2100185, https://doi.org/10.1002/adma.202100185.

[30] F.M. Puglisi, L. Larcher, A. Padovani, P. Pavan, A complete statistical investigation of RTN in HfO2-based RRAM in high resistive state, IEEE Trans. Elect. Dev. 62 (8) (Aug. 2015) 2606–2613, https://doi.org/10.1109/TED.2015.2439812.

[31] S. Vecchi, P. Pavan, F.M. Puglisi, The impact of electrostatic interactions between defects on the characteristics of random telegraph Noise, IEEE Trans. Elect. Dev. 69 (12) (Dec. 2022) 6991–6998, https://doi.org/10.1109/TED.2022.3213502.

[32] O. Gauthier, et al., Enhanced statistical detection of random telegraph noise in frequency and time domain, Solid State Electron. 194 (2022), 108320, https://doi.org/10.1016/j.sse.2022.108320.

[33] B. Hendrickson, R. Widenhorn, P.R. DeStefano, E. Bodegom, Detection and Reconstruction of Random Telegraph Signals Using Machine Learning, arXiv: 2206.00086v1, 31 May 2022.

[34] S.C. Endres, C. Sandrock, W.W. Focke, A simplicial homology algorithm for Lipschitz optimisation, J. Glob. Optim. 72 (2) (2018) 181–217, https://doi.org/10.1007/s10898-018-0645-y.

[35] Thales Becker, et al., An electrical model for trap coupling effects on random telegraph noise, IEEE Elect. Dev. Lett. 41 (10) (2020) 1596–1599.